# Comparison of Sustainable Development Goals Labeling Systems based on Topic Coverage

Li Li*, Yu Zhao** and Zhesi Shen*

*lili2020@mail.las.ac.cn; shenzhs@mail.las.ac.cn
0000-0001-8326-3620; 0000-0001-8414-7912
National Science Library, Chinese Academy of Sciences, China

**zhaoyu.cugb@gmail.com
0009-0006-5554-359X
School of Earth Sciences and Resources, China University of Geosciences, China

**Abstract**
With the growing importance of sustainable development goals (SDGs), various labeling systems have emerged for effective monitoring and evaluation. This study assesses six labeling systems across 1.85 million documents at both paper level and topic level. Our findings indicate that the SDGO and SDSN systems are more aggressive, while systems such as Auckland, Aurora, SIRIS, and Elsevier exhibit significant topic consistency, with similarity scores exceeding 0.75 for most SDGs. However, similarities at the paper level generally fall short, particularly for specific SDGs like SDG 10. We highlight the crucial role of contextual information in keyword-based labeling systems, noting that overlooking context can introduce bias in the retrieval of papers (e.g., variations in "migration" between biomedical and geographical contexts). These results reveal substantial discrepancies among SDG labeling systems, emphasizing the need for improved methodologies to enhance the accuracy and relevance of SDG evaluations.

## 1. Introduction

The Sustainable Development Goals (SDGs) constitute a comprehensive framework of 17 goals established by the United Nations in 2015, aimed at tackling a spectrum of global challenges encompassing social, economic, and environmental dimensions by the target year 2030(Team & Roser, 2023). These goals encompass critical areas such as poverty alleviation, health, education, gender equality, clean energy, climate action, and more.

Since their implementation, scholarly engagement with the SDGs has expanded rapidly, exhibiting a growing diversity in research topics. Nevertheless, the academic consolidation of this field remains incomplete, with notable disparities in research depth across different goals.

Areas such as gender equality (SDG 5)(Eden & Wagstaff, 2021), decent work and economic growth (SDG 8)(Rai et al., 2019), industry, innovation, and infrastructure (SDG 9)(Kynčlová et al., 2020), reduced inequalities (SDG 10)(Oestreich, 2020), life below water (SDG 14)(Recuero Virto, 2018), and peace, justice, and strong institutions (SDG 16)(Hope Sr., 2020).

SDGs have captivated the interest of academia, industry, and policymakers globally. Defining, identifying, and evaluating these goals is essential for aligning development strategies and optimizing resource allocation across institutions, regions, and countries to foster sustainable economic growth. The challenge of achieving these goals transcends national, institutional, and linguistic barriers, underscoring their global relevance. For researchers, accurately pinpointing literature related to SDGs is paramount for the ongoing assessment and refinement of SDGs.

Current methodologies employed to identify SDGs include string matching, machine learning, and text clustering, using a variety of sources such as scientific papers, reports, news, books, Wikipedia, and so on. Research predominantly targets specific goals, like SDG-13 (Climate Action) and SDG-05 (Gender Equality), but lacks a comprehensive, comparative analysis across methods, particularly in research topics, diversity, and relevance of SDG-specific terms within the same dataset(Pradhan et al., 2017).

To bridge these gaps, this study utilizes six major labeling systems to conduct a multidimensional comparison, covering aspects such as research topics, characteristic terms of goals, and system relevance. This analysis provides a scientific basis for refining search strategies in labeling systems. Methodologically, the paper adopts bibliometric approaches and a paper-level topic-label matrix to delve into the diversity and similarity of topics across systems, revealing substantive differences and guiding potential system improvements.

## 2. Literature Review

Research on SDGs has evolved to include various analytical perspectives, such as defining goals, annotating corpora(Meier et al., 2022; Wulff et al., 2023), assessing goals(Giles-Corti et al., 2020), and exploring inter-SDG-relationships(Fonseca et al., 2020; Nilsson et al., 2018; Pradhan et al., 2017; Yang et al., 2020). One prominent focus is the annotation of literature about SDGs, which is pivotal for aligning research with global sustainability objectives.

Studies on SDG labeling systems typically categorize into three approaches: string-matching(Meier et al., 2022), where domain experts develop queries to match texts based on SDG concepts, as seen in systems like Elsevier's(Bedard-Vallee et al., 2023; Jayabalasingham et al., 2019), Auckland's(Wang et al., 2023), SIRIS's(Duran-Silva et al., 2019), Aurora's(Vanderfeesten et al., 2020) and so on; machine learning(Guisiano et al., 2022; Sovrano et al., 2020), which predicts goals through semantic training using models like Word2Vec(Duran-Silva et al., 2019), BERT or unsupervised algorithms(Ferreira et al., 2020); and expert judgment(Yang et al., 2020), where experts directly analyze text content.

However, existing research often focuses on specific fields or individual goals and does not provide a large-scale, systematic comparison across all 17 goals. Moreover, detailed investigations into the semantic features of mislabeling are scarce. This paper leverages extensive data from the scientific literature to compare major labeling systems on coverage, SDG-variation, topic diversity, and system relevance, providing a well-rounded view of their operational efficacy and suggesting pathways for their optimization.

## 3. Research Design

While many studies have analyzed labeling systems, this paper specifically undertakes a systematic comparative analysis of six major labeling systems applied to scientific papers. Due to the unavailability of expert-annotated datasets, this study focuses solely on the outcomes of these specified systems.

*3.1 Research Questions*

Previous studies have rarely addressed the differences in SDG labeling systems across all goals using large samples of scientific literature. To explore how these systems vary in terms of SDG coverage, SDG variation, topic diversity, and similarity, this paper raises the following research questions:

Q1: What is the coverage and variation of the labeling systems regarding the SDGs?
Q2: How similar are the labeling systems at the topic and paper levels?
Q3: If they are not similar, how do these topics distribute and how diverse are these labeling systems?

Q4: What causes the disparity in covered topics across SDG labeling systems for the same goal?

*3.2 Data Acquisition and Processing*

The dataset for this study is derived from the Web of Science Core Collection[1] (SCI-EXPANDED, SSCI) consisting of articles and reviews. The included bibliographic information consists of titles, abstracts, keywords, publication years, and so on. Data were collected in October 2023, and the publication dates of the literature were restricted to the year 2019. After the processes of matching, cleaning, and acquiring citation topics as illustrated in Figure 1, a total of 1.85 million journal articles were obtained. Citation Topics are algorithmically derived citation clusters (using an algorithm developed by CWTS, Leiden)(Traag et al., 2019). This is a three-level hierarchical document-level classification system[2], divided into macro (10 topics), meso (326 topics), and micro (2488 topics), and the information can be accessed through InCites[3]. Similarity is measured using cosine similarity, and the diversity metric employed is the diversity index proposed by (Zhang et al., 2016).

This comprehensive dataset allows for an in-depth analysis of the performance of multiple labeling systems. We used the text2sdg library(Meier et al., 2022) to label the SDGs of our dataset. Text2sdg includes Auckland(version 1), Aurora, SIRIS, Elsevier, SDGO, and SDSN labeling systems. Comparing the similarity of different labeling systems at both topic level and paper level, facilitates the identification of systematic biases or gaps in the coverage and precision of each system, thus contributing significantly to the field of bibliometric research focused on sustainable development goals.

---

[1] https://webofscience.clarivate.cn/wos/woscc/basic-search
[2] https://incites.help.clarivate.com/Content/Research-Areas/citation-topics.htm?Highlight=CITATION%20TOPIC
[3] https://incites.clarivate.com/#/analysis/0/subject

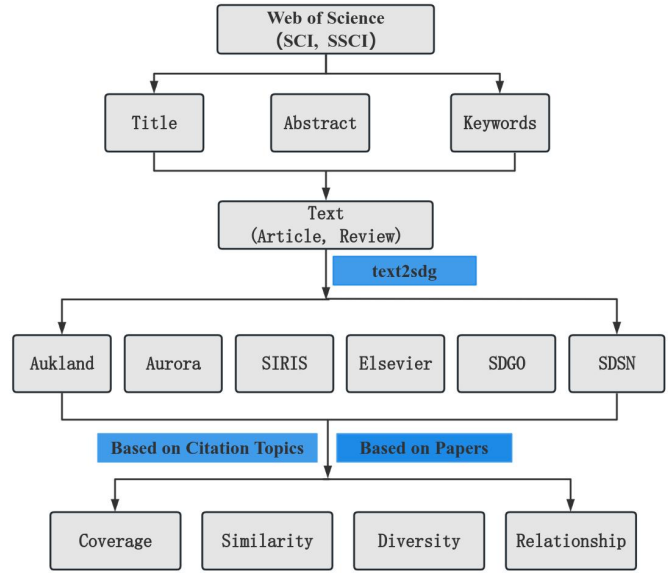

Figure 1: Workflow of data acquisition, clean and analysis.

## 4. Results and Analysis

*4.1 Coverage of Labeling Systems*

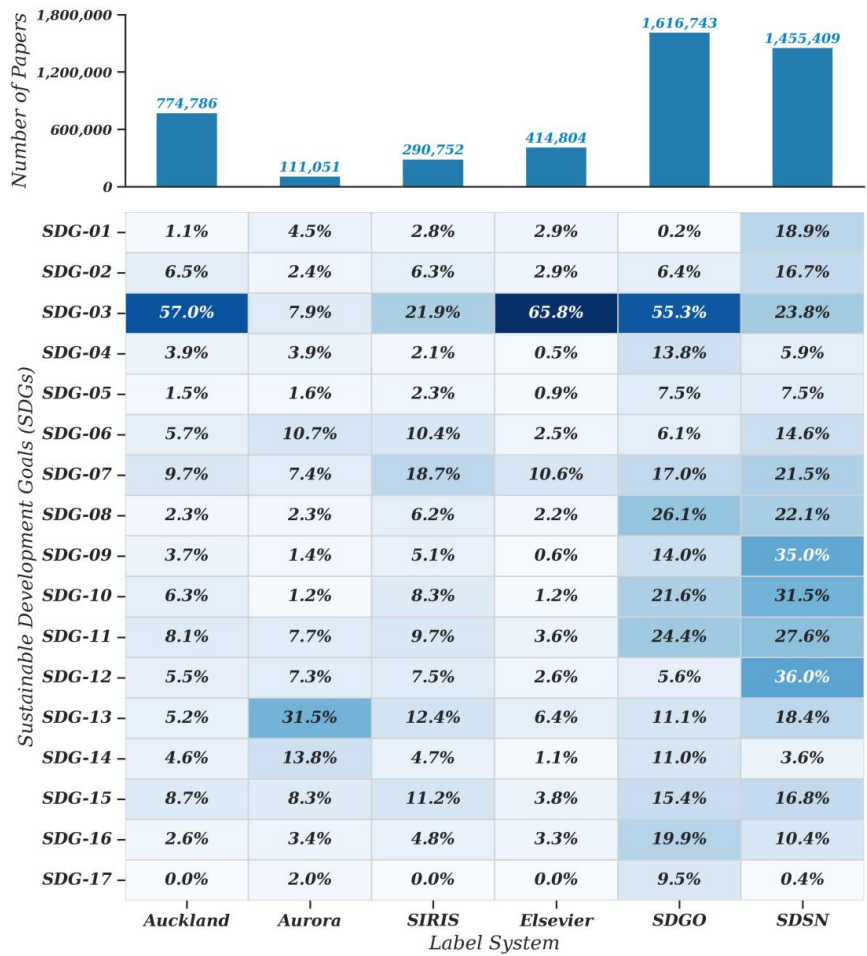

Figure 2: Total Number of Papers Annotated by Different Systems in 2019 and the Proportion of Papers per Goal. Since some papers are annotated with multiple goals, the percentage of papers per SDG may exceed 100%. 2. The bar chart displays the total number of papers annotated by each system.

Figure 2 illustrates the total number of papers annotated by various systems in 2019, as well as the proportion of papers per goal. The data reveals that the SDGO system annotated the most papers, totaling 1.617 million (87.3% coverage), followed by the SDSN system with 1.455 million papers (78.6% coverage). Both systems achieved the highest coverage, annotating over 75% of the papers, significantly surpassing the other four systems—Auckland (41.8%), Elsevier (22.4%), SIRIS (15.7%), and Aurora (6.0%). This indicates that SDGO has the broadest SDG coverage with a more aggressive annotation strategy, whereas Aurora has the narrowest, being relatively conservative.

At the specific SDG, there are notable variations among the systems. Excluding Aurora, the other five systems predominantly favor SDG-3, with Auckland, Elsevier, and SDGO each annotating over 50% of their papers under this goal. Conversely, the Aurora system shows a preference for SDG-13, with more than one-third of its papers focused on this particular goal. This variation in SDGs underscores the different strategic approaches employed by each annotation system, reflecting their operational focus and methodological inclinations.

*4.2 Similarity of Labeling Systems*

The variation of each labeling system for specific goals raised questions about their similarity across these goals. To address this, the study used the Citation Topics information from Incites to create SDGs label-topic matrices and label-paper matrices, calculating the cosine similarity for each goal. The results are displayed in Figures 3 and 4.

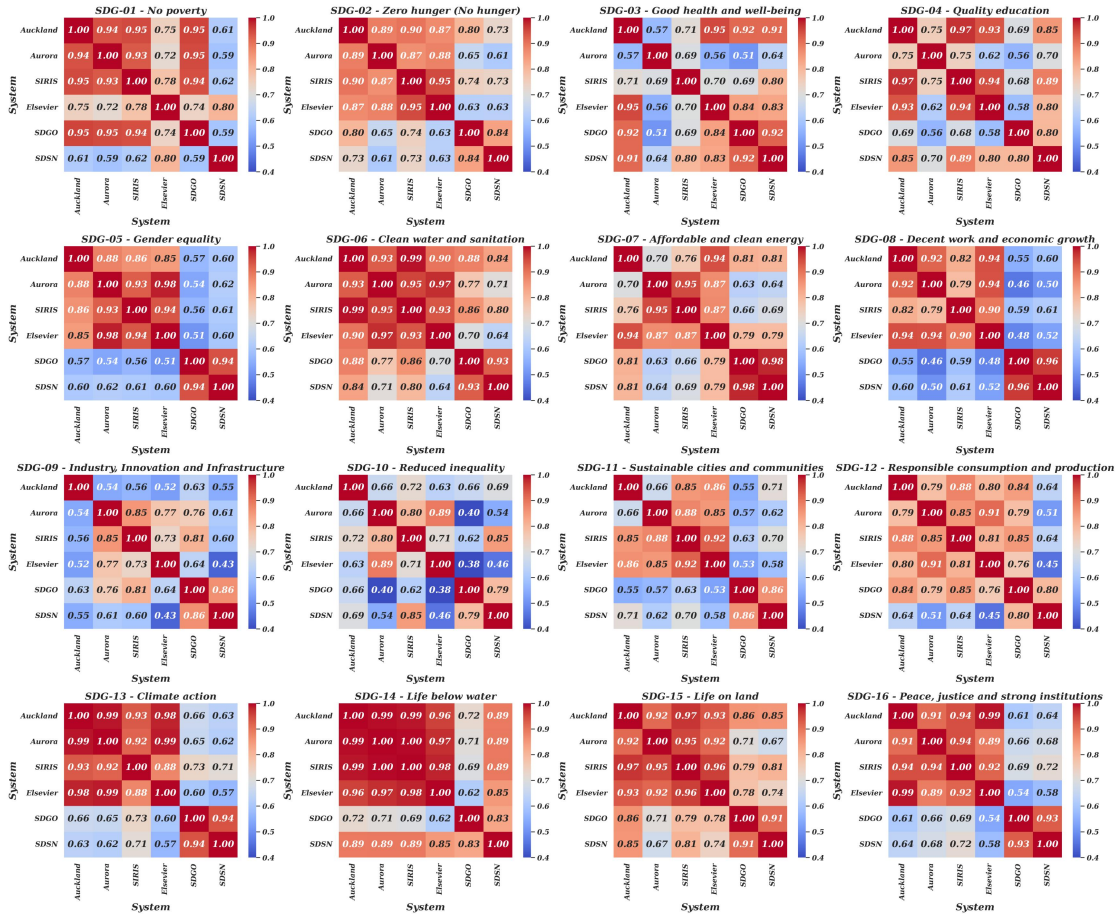

Figure 3: Heatmap of System Similarities Based on Citation Topics. For SDG-17, only the Aurora, SDGO, and SDSN systems identified relevant literature, and thus results for SDG-17 are not shown here.

Figure 3 reveals the Citation Topics similarity indexes among the systems. Auckland, Aurora, SIRIS, and Elsevier show high similarity across several goals, particularly Goals 14, 13, 15, 6, and 16, with similarities exceeding 0.85. However, similarities for Goals 9 and 10 are relatively lower, with Auckland showing less than 0.6 similarity with the other three systems on SDG-9. In contrast, SDGO and SDSN display high similarity, especially on Goals 7 and 8, where similarity exceeds 0.95.

This indicates that on a topic level, Auckland, Aurora, SIRIS, and Elsevier share similarities, as do SDGO and SDSN. However, there are differences in specific goals such as Goals 3, 9, and 10, likely influenced by the respective systems' search strategies.

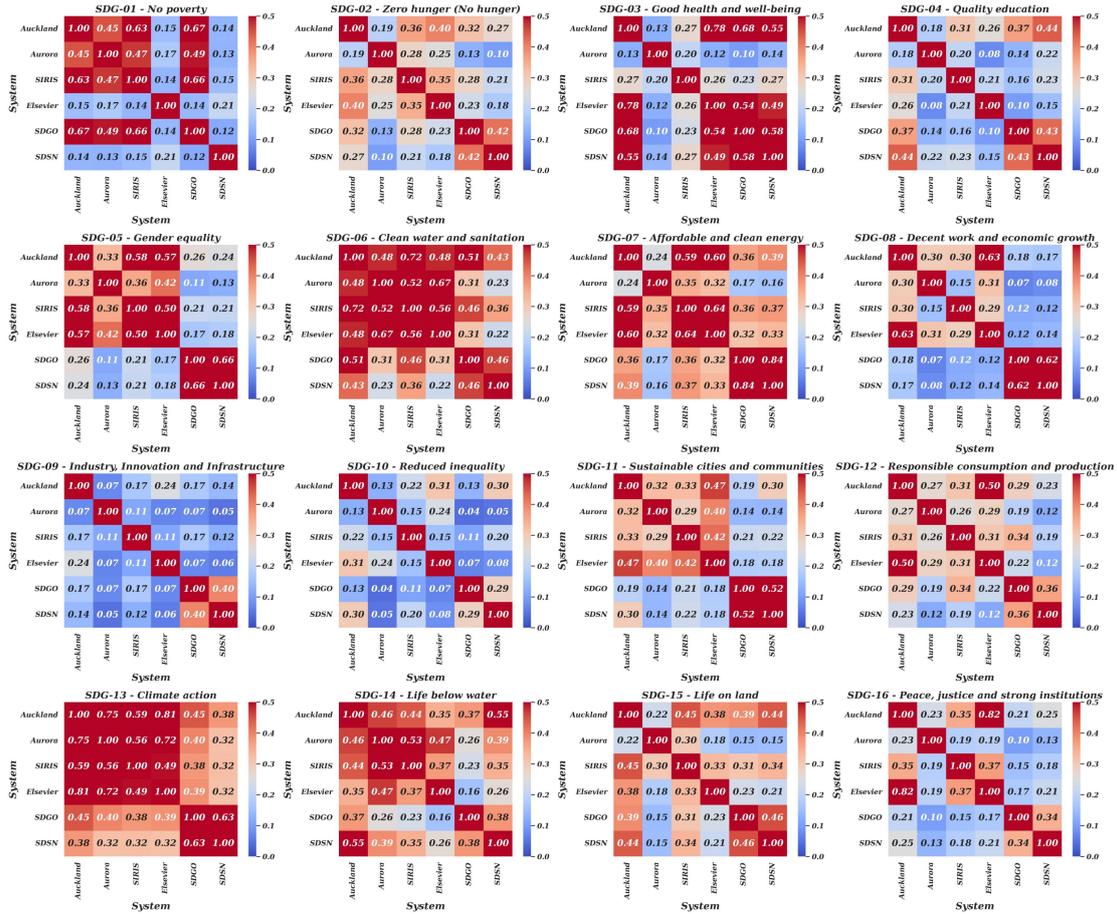

Figure 4: Heatmap of System Similarities Based on Papers. The color bar is set from 0 to 0.5 to highlight differences in similarity.

Figure 4 assesses the paper-based similarities across systems. Overall, similarities at the paper level are generally low, mostly below 0.5. Compared to Figure 3, the paper-based similarities reflect similar patterns, with Auckland, Aurora, SIRIS, and Elsevier showing comparable levels, and SDGO and SDSN also displaying similarities. Notably, for SDG-10, the similarity among systems is particularly low, not exceeding 0.3, indicating significant divergence in how systems identify SDG-10, with most articles being annotated by only one system.

These findings illustrate both topic and paper-level patterns of similarity and divergence among the labeling systems, highlighting the complex dynamics of how different systems approach the annotation of SDGs. This analysis is crucial for understanding the strengths and limitations of each system, potentially guiding enhancements in SDG annotation strategies.

*4.3 Diversity of Topics in Labeling Systems*

Following the analysis of system similarities in section 4.2, which revealed varying degrees of similarity across different goals, this section delves further into the diversity and distribution of topics within each system. To measure the topic diversity of each annotation system, the study employed the diversity index(Zhang et al., 2016). The results are presented in Table 1.

**Table 1: Topic Diversity Index for Each System Across Goals**

| SDGs | Auckland | Aurora | Elsevier | SDGO | SDSN | SIRIS | Max-Min |
|---|---|---|---|---|---|---|---|
| SDG-01 | 2.30 | 2.34 | 3.26 | 2.26 | 4.05 | 2.39 | 1.79 |
| SDG-02 | 3.13 | 2.63 | 2.62 | 3.34 | 4.08 | 2.95 | 1.46 |
| SDG-03 | 3.04 | 2.77 | 2.98 | 3.76 | 3.38 | 3.51 | 0.98 |
| SDG-04 | 2.13 | 2.04 | 1.79 | 3.72 | 2.68 | 1.92 | 1.94 |
| SDG-05 | 2.02 | 1.81 | 1.73 | 2.99 | 3.14 | 1.82 | 1.41 |
| SDG-06 | 2.69 | 2.47 | 2.31 | 3.40 | 3.60 | 2.55 | 1.29 |
| SDG-07 | 2.22 | 2.37 | 2.36 | 3.27 | 3.32 | 2.58 | 1.10 |
| SDG-08 | 2.00 | 1.80 | 1.84 | 4.22 | 4.21 | 2.66 | 2.42 |
| SDG-09 | 3.41 | 2.56 | 1.91 | 4.01 | 4.13 | 2.66 | 2.22 |
| **SDG-10** | 3.35 | 2.00 | **1.91** | **4.27** | 3.65 | 3.05 | **2.36** |
| SDG-11 | 3.09 | 2.17 | 2.40 | 3.96 | 4.09 | 2.64 | 1.92 |
| SDG-12 | 2.94 | 2.64 | 2.61 | 3.88 | 4.06 | 3.13 | 1.45 |
| SDG-13 | 2.22 | 2.24 | 1.96 | 3.46 | 3.27 | 2.52 | 1.50 |

| SDG-14 | 2.36 | 2.11 | 1.49 | 3.35 | 2.55 | 2.01 | 1.86 |
| SDG-15 | 2.41 | 2.05 | 1.85 | 3.60 | 3.40 | 1.94 | 1.76 |
| SDG-16 | 2.04 | 1.98 | 1.88 | 3.57 | 3.28 | 2.25 | 1.69 |
| Average | 2.58 | 2.25 | 2.18 | **3.57** | 3.55 | 2.53 | **1.70** |

Note: The topic classification used is the meso-level of Citation Topics.

Table 1 shows that the average topic diversity indices range from 1.70 to 3.57, with SDGO exhibiting the highest average diversity, indicating a broad distribution of identified topics. Notably, SDG-10 displays the most significant differences in topic diversity among the systems, with SDGO achieving the highest diversity score of 4.27 and Elsevier the lowest at 1.91.

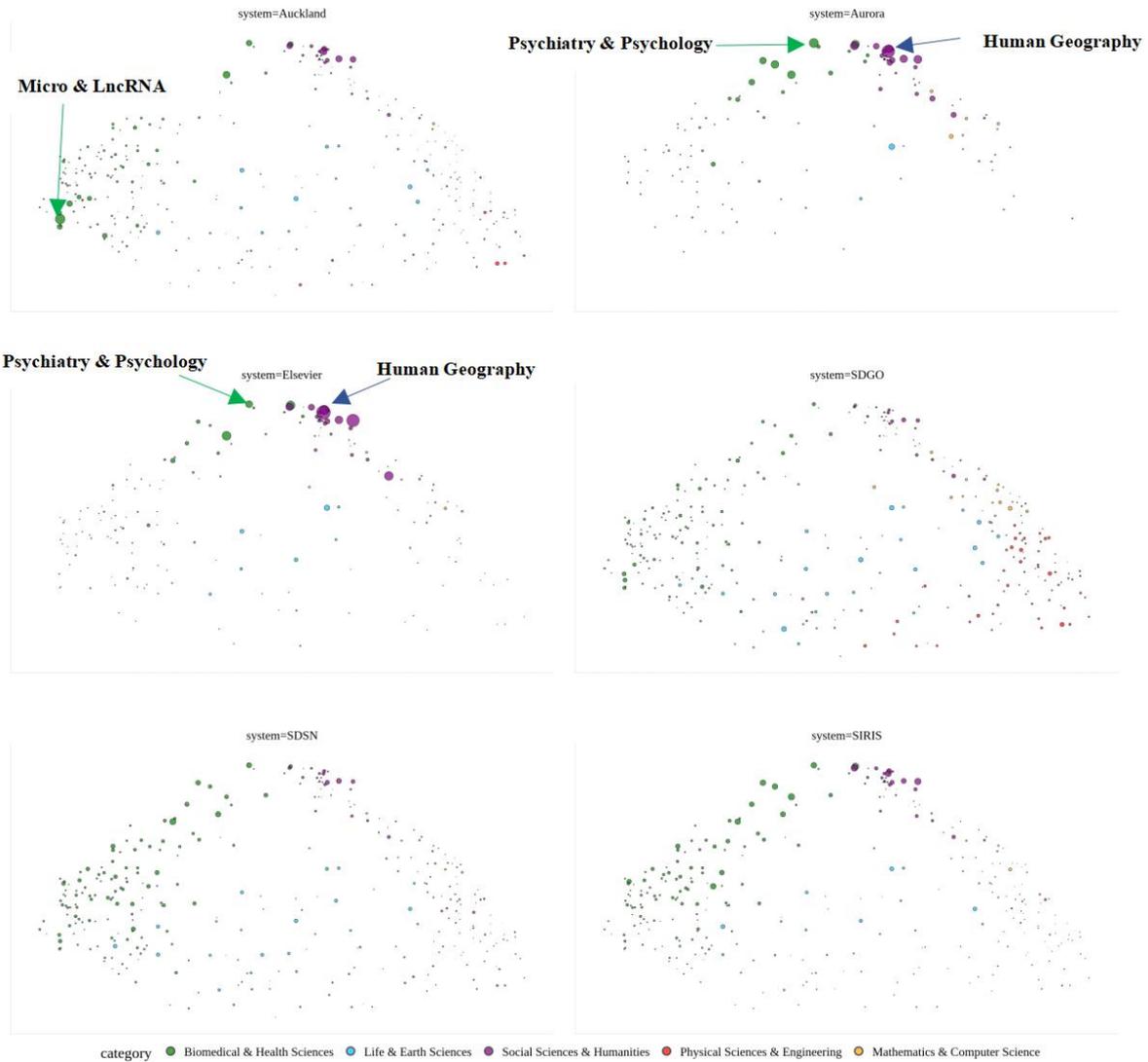

**Figure 5: Topic Distribution of Papers under SDG-10 for Each System.** Each dot reprents a Citation Topic with size representing the fraction of retreved papers and color representing macro-level fields. The distance among dots represent the similarity among citation topics, with more similar topics locating more closely. The layout is generated based on the citation networks of citation topics(Zhang & Shen, 2024).

Figure 5 illustrates the topic distribution of papers under SDG-10. Aurora and Elsevier systems have a narrower range of topics, while SDGO, SDSN, Auckland, and SIRIS cover a broader spectrum, accounting for over 90% of the research topic (300 out of 326). Aurora and Elsevier's topics are relatively concentrated in Social Sciences & Humanities and Biomedical & Health Sciences. In Social Sciences & Humanities, the predominant topic is Human Geography, focusing on spatial inequalities, social exclusion, and resource distribution, as well as promoting social inclusion and participation. In Biomedical & Health Sciences, the primary focus is Psychiatry & Psychology, emphasizing the enhancement of mental health services accessibility, addressing mental health disparities, advocating for mental health rights and inclusiveness, and promoting mental health education and prevention.

One noteworthy finding is the "Micro & Long Noncoding RNA" topic in Auckland, which encompasses 3,330 papers, representing 6.8% of the total, significantly higher than other topics. Detailed data for this finding is provided in Appendix Table 1. This detailed exploration into topic diversity and specific concentrations provides a deeper understanding of how different labeling systems approach SDG-related research, highlighting the varying scopes and focuses that influence their effectiveness in capturing the multifaceted aspects of sustainable development goals.

*4.4 SDG-Specific Keyword Relationship Networks*

The distribution of topics for an SDG reveals that identical keywords may have different meanings across various domains. Some labeling systems fail to fully capture the semantic information of keywords under different contexts, leading to annotation inaccuracies. Therefore, this study constructed co-occurrence networks of characteristic keywords for specific goals to reveal their semantic nuances and enhance the accuracy of labeling systems. The analysis focused on SDG-10, particularly examining the topic "Micro & Long Noncoding RNA" and "Human Geography" in the Auckland system, highlighting the keyword "Migration" within their co-occurrence networks as shown in Figure 6.

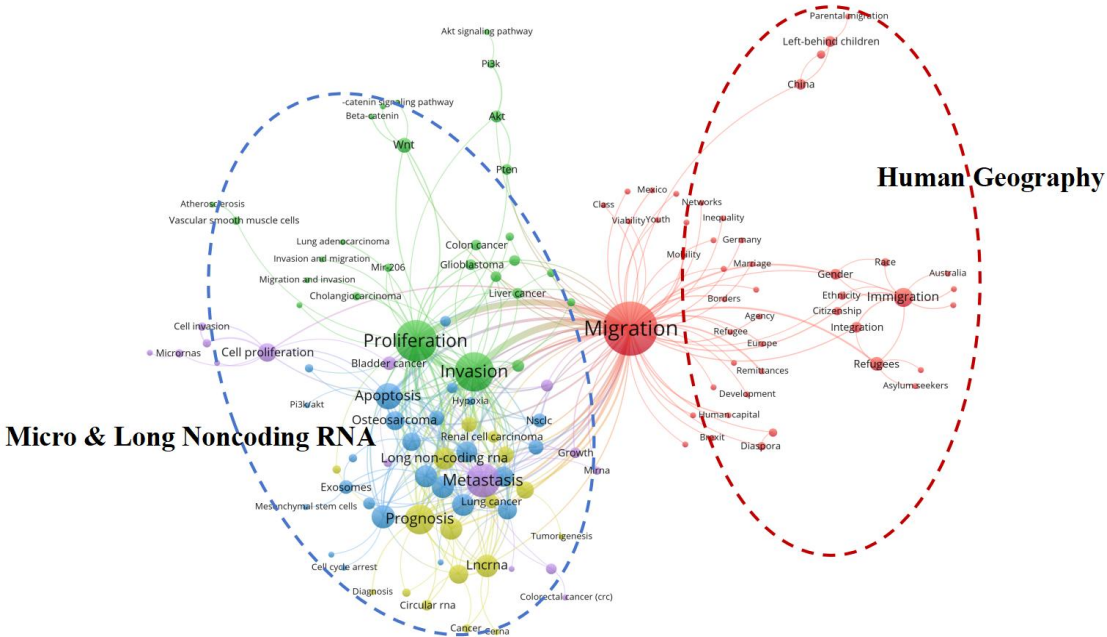

**Figure 6: Co-occurrence Network for SDG-10 in the Auckland System.** The figure was generated using VOSviewer(van Eck & Waltman, 2010), and only the co-occurrences with a frequency of five or more times are displayed.

The figure demonstrates how the keyword "Migration" carries different meanings in various contexts. For instance, under "Micro & Long Noncoding RNA" it relates to cancer cell metastasis and genetic material expression, which diverges significantly from the interpretations relevant to SDG-10. Conversely, "Human Geography" pertains to gender, racial, and regional variations among migrants, aligning closely with the topic of SDG-10. This analysis underscores the necessity of considering the semantic implications of keywords when annotating targets.

## 5. Conclusions and Discussion

*5.1 Conclusions*

This paper systematically compared different SDG labeling systems at the paper level and topic level, focusing on their coverage, topic and paper similarities, topic diversity, and keyword co-occurrence networks. Key findings include:

(1) SDGO and SDSN annotated papers more aggressively, exceeding 75% coverage, while Aurora was more conservative, focusing mainly on SDG-13 compared to other systems that concentrated on SDG-3.
(2) Auckland, Aurora, SIRIS, and Elsevier showed higher topic and paper similarities, particularly above 0.75 for most goals at the topic level, with lower paper-level similarities generally under 0.5.
(3) SDGO exhibited generally higher topic diversity, covering a broader range of topics across multiple goals, whereas Elsevier's topics were more concentrated and less diverse.
(4) The same keyword can have entirely different semantic implications depending on the context, as seen in Auckland's system where "Migration" varied dramatically between biomedical and geographical topics.

*5.2 Discussion*

This study compared the annotation performances of different systems, revealing their focus on distinct SDGs and the critical need to consider keyword semantics during annotation. Limitations of this study include:

(1) The data set was restricted to 2019 papers, and updates to labeling systems could impact results and lead to mislabeling.
(2) The selected systems predominantly used designated keywords for annotation, excluding systems that utilize machine learning and semantic text analysis, which limits the comparative analysis to specific dimensions and scopes.
(3) The similarity was only analyzed at the topic and paper levels, not semantically, which could have revealed finer distinctions between systems.
(4) The absence of a "gold standard" annotation system limited the ability to comprehensively identify strengths and weaknesses across systems.

Future research could incorporate these factors and include different types of labeling systems, such as those utilizing machine learning and semantic analysis, to help establish a "gold standard" for system evaluation.

**Open science practices**

Due to the nature of the research data used in this paper, which includes bibliographic information such as titles, abstracts, and keywords, access to the Web of Science (WoS) database is restricted by a contract with Clarivate that prohibits the redistribution of their data. Researchers interested in conducting their analysis with the raw data will need to obtain it through a paid subscription to Clarivate.

**Acknowledgments**

This paper utilizes GPT technology for language polishing of the manuscript.

**Author contributions**

**Li Li**: Writing – original draft, Formal analysis, Data curation, Methodology, Visualization, Conceptualization.

**Yu Zhao**: Writing – review & editing, Formal analysis, Conceptualization.

**Zhesi Shen**: Conceptualization, Writing – review & editing, Formal analysis, Supervision, Methodology .

**Competing interests**

The authors declare that they have no conflict of interest.

**Appendix**

Table 1: Top 5 Citation Topics for SDG-10 Across Different Labeling Systems

| Meso topics | Category | Auckland | Aurora | Elsevier | SDGO | SDSN | SIRIS |
|---|---|---|---|---|---|---|---|
| Human Geography | SSH | 1845 (3.8%) | 163 (11.7%) | 686 (13.8%) | - | - | 686 (2.9%) |
| Economics | SSH | - | 66 (4.8%) | 579 (11.6%) | - | - | 706 (2.9%) |
| Political Science | SSH | - | - | 316 (6.3%) | - | - | - |
| Sustainability Science | SSH | - | - | 286 (5.7%) | - | - | - |
| Healthcare Policy | BHS | 1743 (3.6%) | - | 280 (5.6%) | - | - | 747 (3.1%) |
| Nursing | BHS | - | - | - | - | 9701 (2.1%) | - |
| Micro & Long Noncoding RNA | BHS | 3330 (6.8%) | - | - | - | - | - |
| Psychiatry | BHS | - | - | - | - | 7583 (1.7%) | - |
| Nutrition & Dietetics | BHS | - | - | - | - | 12560 (2.7%) | - |
| Forestry | LES | - | - | - | 5214 (1.5%) | - | - |
| Crop Science | LES | - | - | - | 5793 (1.7%) | - | - |
| Marine Biology | LES | - | - | - | 5539 (1.6%) | - | - |
| Electrochemistry | PSE | - | - | - | 4747 (1.4%) | - | - |
| Psychiatry & Psychology | BHS | - | 84 (6.1%) | - | - | 9008 (2%) | - |
| Gender & Sexuality Studies | SSH | 1296 (2.7%) | 65 (4.7%) | - | - | - | 765 (3.2%) |
| Management | SSH | 1447 (3%) | - | - | 4507 (1.3%) | 8346 (1.8%) | - |
| Social Psychology | BHS | - | 67 (4.8%) | - | - | - | 780 (3.3%) |

Notes: 1. The numbers before the parentheses indicate the number of papers on that topic; the numbers inside the parentheses represent the percentage of papers on that topic relative to the total number of papers on the topic within the system. 2. Abbreviations used are SSH for Social Sciences & Humanities, BHS for Biomedical & Health Sciences, LES for Life & Earth Sciences, and PSE for Physical Sciences & Engineering.